\documentclass[conference]{IEEEtran}
\IEEEoverridecommandlockouts
\usepackage{cite}
\usepackage{amsmath,amssymb,amsfonts}
\usepackage{algorithmic}
\usepackage{graphicx}
\usepackage{textcomp}
\usepackage{xcolor}
\def\BibTeX{{\rm B\kern-.05em{\sc i\kern-.025em b}\kern-.08em
    T\kern-.1667em\lower.7ex\hbox{E}\kern-.125emX}}
\begin{document}

\title{Speaker Placement Agnosticism: Improving the Distance-based Amplitude Panning Algorithm\\
\thanks{Thanks to Shahrokh Yadegari (Calit2) and Jos\'{e} Miguel Fernandez (IRCAM) for providing invaluable listening feedback during development.}
}

\author{\IEEEauthorblockN{Jacob Sundstrom}
\IEEEauthorblockA{\textit{Department of Music} \\
\textit{University of California, San Diego}\\
La Jolla, California, USA \\
jsundst@ucsd.edu}
% \IEEEauthorblockA{\textit{Department of Music} \\
% \textit{University of California, San Diego}\\
% La Jolla, California, USA}
% \IEEEauthorblockA{\textit{AIXDSP, Inc.}\\
% Scappoose, Oregon, USA \\
% jacobsundstrom@gmail.com}
% \and
}

\maketitle

\begin{abstract}
Lossius et. al introduced the distance-based amplitude panning algorithm, or DBAP, to enable flexibility of loudspeaker placement in artistic and scientific contexts. The algorithm allows for arbitrary loudspeaker locations in a 2D plane so that a virtual sound source may navigate the 2D space. The gains for each speaker are calculated as a function of the source's distance to each loudspeaker, thus creating a sound field. This gives the listener the impression of a source moving through the field of loudspeakers. This paper introduces a heuristically developed robust variation of DBAP that corrects for faulty assumptions in the implementation of Lossius. Specifically, this paper develops a method for working with sound sources outside the field of loudspeakers in which the Lossius version produces distorted aural impressions and wildly undulating amplitudes caused by spatial discontinuities in the gains of the various loudspeakers. In smoothing the spatial impression of the virtual source, we are also able to eliminate the calculation of the convex hull entirely, a necessary component of the original implementation. This significantly simplifies and reduces the calculations required for any space in either two or three dimensions.
\end{abstract}


\section{Introduction}
Spatialized sound generally works from a set of loudspeakers that are placed along the permiter of a ring (2D) or the surface of a sphere (3D), evenly enclosing a listening area. There are several well-known paradigms which use this loudspeaker configuration, most notably vector-based amplitude panning (VBAP) \cite{vbap} and Ambisonics \cite{ambi}. Few paradigms have been developed for arbitrary placement of loudspeakers. Ville Pulkki, who also created VBAP, developed multiple-direction amplitude panning \cite{mdap} and in 2009, Lossius, Baltazar, and de la Hogue published the technique known as distance-based amplitude panning (DBAP) \cite{dbap} upon which the method in this paper was developed. DBAP currently has implementations in a number of computer music softwares such as Pd \cite{pd_dbap}, Max \cite{max_dbap}, Jamoma \cite{jamoma_dbap}, and SuperCollider \cite{sc_dbap}.

There are multiple aesthetic circumstances that would stand to benefit from a DBAP-like paradigm of a flexible loudspeaker layout. An algorithm which compensates for arbitrarily positioned loudspeakers would enable new aesthetic experiences with spatialized sound, as well as improve the physical layout of artistic installations which utilize spatial sound. Without a flexible loudspeaker layout, loudspeakers for spatialized sound must be constrained to a ring or sphere. Any alternative layout, say in an oval in a rectangular room, will result in distortion of the sound field.

In addition, use of so-called flocking algorithms such as Craig Reynolds' Boids algorithm \cite{boids} are particularly well-suited to a loudspeaker field since each agent in a flock is given an (x,y) or (x,y,z) coordinate that can be mapped into the ``loudspeaker-space'' without conversions or other compensation.

\section{Distance-based Amplitude Panning}
As noted, distance-based ampltude panning (DBAP) is a sound spatialization paradigm which allows a virtual source to be placed in a field of arbitararily placed loudspeakers by calculating the gain of each loudspeaker as a function of the distance to the virtual source. The gain, $v_i$, for the $i$th loudspeaker is:
\begin{equation}
    v_i = \frac{k w_i}{d^a_i} \label{vi_orig}
\end{equation}
where
\begin{equation}
    k = \frac{1}{\sqrt{\sum_{i=1}^{N} \frac{w_i^2}{d_i^{2a}}}} \label{k_orig}
\end{equation}
\begin{equation}
    a = \frac{R}{20\log_{10} 2}
\end{equation}
and $N$ is the number of loudspeakers, $d_i$ is the distance from the source to the $i$th loudspeaker, $w_i$ is a weighting paramter for the $i$th loudspeaker typically set to 1, and $a$ is a coefficient calculated from a rolloff, $R$, in decibels. The variable $k$ is a coefficient that is a function of the position of the source and all the speakers.

In two dimensions, the distance $d_i$ between the virtual source position, $(x_s, y_s)$, and the position of the $i$th loudspeaker, $(x_i, y_i)$, is defined as:
\begin{equation}
    d_i = \sqrt{ (x_i - x_s)^2 + (y_i - y_s)^2 + r_{s}^2} \label{distance}
\end{equation}
where $r_s$ is a spatial blur factor.\footnote{In \cite{dbap}, Lossius constrains $r_s \geq 0$. This is unnecessary since $r_s$ is squared in \eqref{distance}.}

In doing so, DBAP enables spatial sound without the normal constraints of loudspeaker placement at the expense of loss of clarity in spatial position for sparsely populated loudspeaker fields. Note that this approach is easily apadted to three dimensions by adding the additional dimension in \eqref{distance}.
% d_{i} = \sqrt{(x_i - x_s)^2 + (y_i - y_s)^2 + r_s^2} \label{dbap_orig}

\section{Difficulties with DBAP}
Unfortuntely, DBAP as presented in \cite{dbap} is error prone. It is subject to spatial distortions in specific but relatively common situations and configurations. The primary difficulty in the implementation is how to approach the problem of a virtual source positioned outside the field of loudspeakers.

In \S 2.6 of \cite{dbap}, the issue of sources outside the field of speakers is presented. As the virtual source moves further and further away, the difference between the gains for each speaker is reduced (i.e. the ratio of each distance to every other one approaches 1). This causes the source to appear to move towards the center of the loudspeaker field as it fades. To counteract this, the proposed solution is that the user project the source that is outside the field onto the convex hull of the loudspeaker field --- where the projection is defined as the point on the hull that is closest to the source outside --- and to use this projected point in subsequent calculations of gain. This biases the gains of all the speakers in the direction of the virtual source. Projecting would likewise give the distance from the source to the convex hull and thus allow the amplitudes to be scaled as a function of the distance from the hull, typically using the inverse square law: $1/d^2$. In the DBAP algorithm, this can be modified to $1/d^{2a}$ to account for the value $a$. The projection provides sufficient biasing of the gains to maintain the spatial illusion, although depending on the scaling function, significant distortions may occur. This presents two difficulties: the calculation of the convex hull and nonunique solutions for speaker gains when projecting onto the convex hull.

\begin{figure}
\centerline{\includegraphics[width=0.5\textwidth]{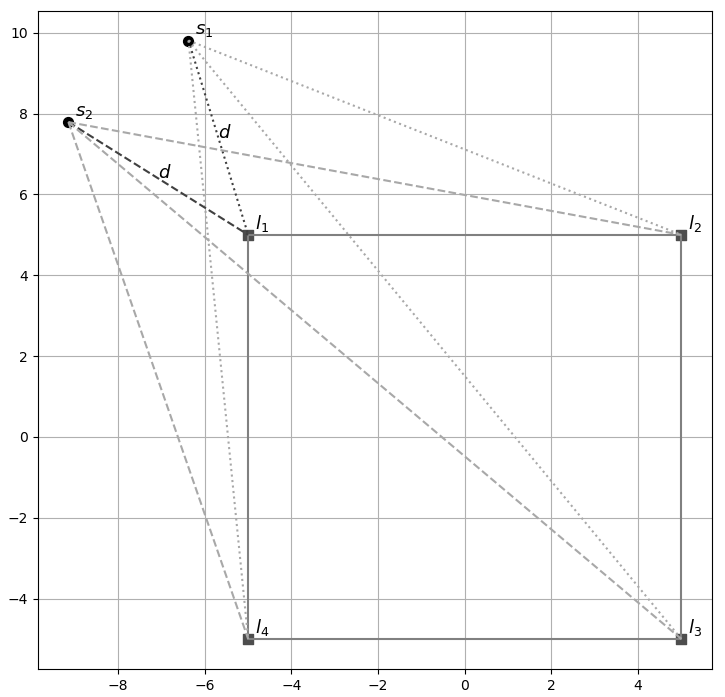}}
\caption{Both $s_1$ and $s_2$ have the same projection, $l_1$, and are the same distance, $d$, from $l_1$. This results in identical gain calculations and therefore, no spatial differentiation. Note, though, that the distance to $l_2$, $l_3$, and $l_4$ are different. When using the original DBAP algorithm, the solutions for $s_1$ and $s_2$ are identical and they are perceptably undifferentiable in space.}
\label{circlesource}
\end{figure}

\subsection{Calculation of a Convex Hull}
Algorithms for the calculation of a convex hull are well-documented in computer science literature (\cite{grahamScan}, \cite{jarvisMarch}, \cite{quickhull}). However, adapting these algorithms for three-dimensions is not trivial and often highly complex. Even after the calculation of a convex hull it must be determined, at each time point, whether or not the virtual source is contained within the hull. If it is not contained within the convex hull, one must then calculate the orthogonal projection of the source onto the hull which generally involves solving a quadratic programming problem, a resource-intensive process in two dimensions becoming even more so in three dimensions \cite{gilbertProjection}.

\subsection{Nonunique Solutions in the Projection Process}
\begin{figure}
% \centerline{\includegraphics[width=0.5\textwidth]{figs/quad/orig_gain.png}}
\centerline{\includegraphics[width=0.5\textwidth]{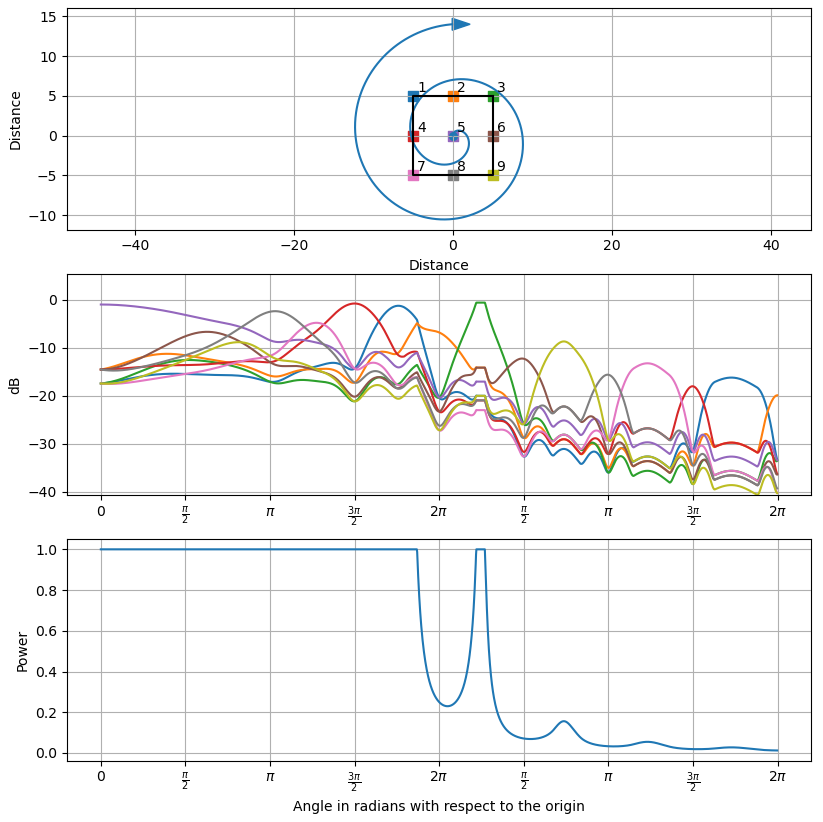}}
\caption{Original DBAP. Rolloff $= 6$, $r_s = 1.073$. \emph{Top}: Plot of a 3x3 grid of loudspeakers with a line showing the movement of the source from the center of the field, extending in a spiral to a point outside the field. \emph{Middle}: Gains of the each speaker, color coded to correspond to the above figure. Note the flat spot in the third speaker (speaker 2) where the percieved motion stalls due to the nonunique projection. \emph{Bottom}: The power of the entire field as a function of the source's angle.. Notice the undulating power as the source traverses the boundary of the convex hull.}
\label{original_gain_plots}
\end{figure}

In cases where the projection is orthogonal to the perimeter of the convex hull, this method of projection to create biasing works adequately. However, when a projection is not orthogonal, as in the case of the area beyond the vertices of the convex hull where the projection is equal to one of the vertices, this method fails to provide a unique solution. In particular, all spatial differentiation is lost when different sources share the same projection which occurs when the projection is the same vertex (Fig. \ref{circlesource}) and that vertex is then used in all subsequent calculations of gain. In the most obvious case, if a source were to move in a circle around a loudspeaker which is also a vertex of the convex hull, there will be an area in which the source will appear not to move, since the projected distance \emph{and} the point of projection are identical, even though the source's relative distance to every other speaker is different (Figure \ref{circlesource}). This is illustrated in the middle plot of Figure \ref{original_gain_plots}, one can see the flat spot in the third speaker just past the $2\pi$ mark where the apparent movement of the source stops. Indeed, this phenomenon occurs in every other place where the projection is a vertex and the change in amplitude in those instances is solely a function of the source's increasing distance from the hull. This can be seen with the fixed gain relationships between all speakers. This effect is exacerbated as $r_s$ is reduced and the rolloff $R$ is decreased.

Additionally, a source crossing the threshold of the convex hull can result in spatial discontinuities when utilizing this method. By decreasing the gain of the virtual source by $1/(d+1)^2$ where $d$ is the distance from the source to the hull, the power begins to undulate wildly. This is readily seen in the bottom plot of Figure \ref{original_gain_plots} at just before the $2\pi$ mark in the center of the plot. Although this example is presented in two dimensions, it is easy to see how it would apply in the same way in three dimensions and therefore is unnecessary to work out.

\subsection{Absolute Distance-based Amplitude Panning}
The authors in \cite{coreography} found that the implementation of DBAP as proposed in \cite{dbap} produced insufficient clarity and that they ``discovered that especially the trajectory of moving sounds ... appears more clearly shaped or ``sharper'', compared to the unmodified DBAP algorithm.'' Moreover, the authors required the source to venture outside the loudspeaker field (convex hull) and it appears that they na{\"i}vely implemented DBAP without regard for what was noted in \S 4.6 of \cite{dbap}. Thus, they had a need to not only clarify the spatial trajectories but also to solve the issue of a source outside the hull.

The authors instead propose to drop the constant intensity condition in the original implementation. This resulted in what they termed absolute distance-based amplitude panning, or ADBAP. While the authors did not disclose the exact calculation of gain they used, it can be inferred that they simply set $k = 1$ for all source positions. This results in the gain, $v_i$, calculated as
\begin{equation}
    v_i = \frac{w_i}{d_i^a} \label{adbap}
\end{equation}

While ADBAP provides superior imaging as noted by the authors of \cite{coreography}, the total power from all loudspeakers in constantly in flux. In a non-grid layout, ADBAP is prone to large changes in power when a source passes through a cluster of speakers. In the interest of completeness of comparison, plots of the behavior of the ADBAP algorithm are presented in Figure \ref{abs_gain_plots} alongside the original version as well as the modified version here.

\section{New Methods}\label{new_methods}
In contrast to the original DBAP algorithm, the proposed methods were developed less from a theoretical framework and instead built upon an existing algorithm using an empirical and heuristically driven process, although the constant intensity condition remains. Thus the proposed solutions are not guarenteed to be perfect but have, in practice, been shown to provide a more convincing spatial impression given the placement and movement of a virtual source.

\subsection{Choice of the Loudspeaker Layout}
\begin{figure}
% \centerline{\includegraphics[width=0.5\textwidth]{figs/quad/abs_gain.png}}
\centerline{\includegraphics[width=0.5\textwidth]{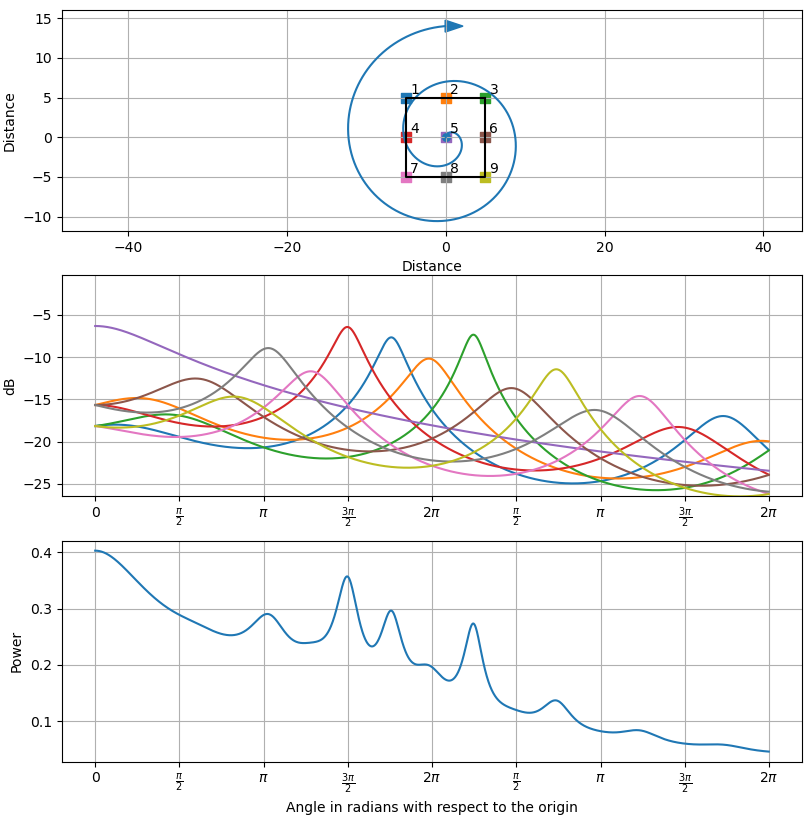}}
\caption{ADBAP. Rolloff $= 6$. \emph{Top}: Plot of a 3x3 grid of loudspeakers with a line showing the movement of the source from the center of the field, extending in a spiral to a point outside the field. \emph{Middle}: Gains of the each speaker, color coded to correspond to the above figure. \emph{Bottom}: The power of the entire field as a function of the source's angle.}
\label{abs_gain_plots}
\end{figure}

To illustrate each difficulty and their proposed solutions, we will analyze a simple 3x3 grid of 9 speakers (Figures \ref{original_gain_plots}, \ref{new_gain_plots}, and \ref{abs_gain_plots}, top) where the convex hull is a square with a loudspeaker at each vertex. Although unconventional, a grid of speakers is the most ideal loudspeaker layout for all forms of DBAP since it spatially samples the area of the loudspeaker field evenly in all dimensions. The source will move outward in a spiral from the origin clockwise until it is placed outside the convex hull of loudspeakers. This allows us to examine the behavior of the various algorithms when a source is moving inside the hull as well as the behavior when it traverses the boundary of the hull and moves outside it. This particular layout was auditioned and proved to be the most useful with regard to analysis and comparison.\footnote{A note about the figures: The many different and overlapping gain lines make the greyscale plots, which are necessary for publication, difficult to read. Therefore, larger, color, and easier to read plots are available on the authors website at http://jacobsundstrom.com/research/dbap.}

Also included are plots for a random layout of 10 loudspeakers; this is presented in the Appendix. Finally, a set of plots is provided on the authors website that illustrate DBAPs behavior in a quadraphonic and 9-speaker circular (nonagon) layout. It should be noted that DBAP, in any of its flavors, ought \emph{not} be the first choice for spatialzation algorithms in any regular polygonal layout. DBAP is specifically designed for \emph{arbitrary loudspeaker positions} which would cause other algorithms, notably VBAP or Ambisonics, to fail. An additional set of figures for all plots are included which plot the SPL a listener would experience as a function of both the angle of the virtual source and the ''incoming angle" on a heatmap. That is, the x-axis is the angle of the virtual source in radians and the y-axis is the ''incoming angle" in radians where 0 is directly ahead and $\pi$ is directly behind a listener facing forward (0 radians). These plots assume all speakers are facing the listener and the SPL is scaled as $1/d^2$ for all speakers.\footnote{The SPL heatmaps are not included in the paper since they need to be large in order to be adequately read and printing them in grayscale exacerbates this problem.}

\subsection{Introduction of the variable $p$}
In order to more adequately and smoothly scale the gains of a source moving in and out of the field of speakers, we drop the assumption that the powers must always equal 1 and that when a source travels outside of the convex hull, the amplitude of the source must be scaled. Instead we introduce the variable $p$ in the calculation of $k$ where

\begin{equation}
    k = \frac{p^{2a}}{\sqrt{\sum_{i=1}^{N} \frac{w_i^2}{d_i^{2a}}}} \label{k_new}
\end{equation}
and
\begin{equation}
    p = \begin{cases} q = \frac{ \max{(d_{s})} }{ d_{rs} }, & \mbox{if } q\mbox{ $<$ 1} \\ 1, & \mbox{otherwise} \end{cases}
    % p = \begin{cases} q = \frac{ \max\{d_1, ... , d_{N} \} }{ d_{ir} }, & \mbox{if } q\mbox{ $<$ 1} \\ 1, & \mbox{otherwise} \end{cases}
\end{equation}

\begin{figure}
% \centerline{\includegraphics[width=0.5\textwidth]{figs/quad/new_gain.png}}
\centerline{\includegraphics[width=0.5\textwidth]{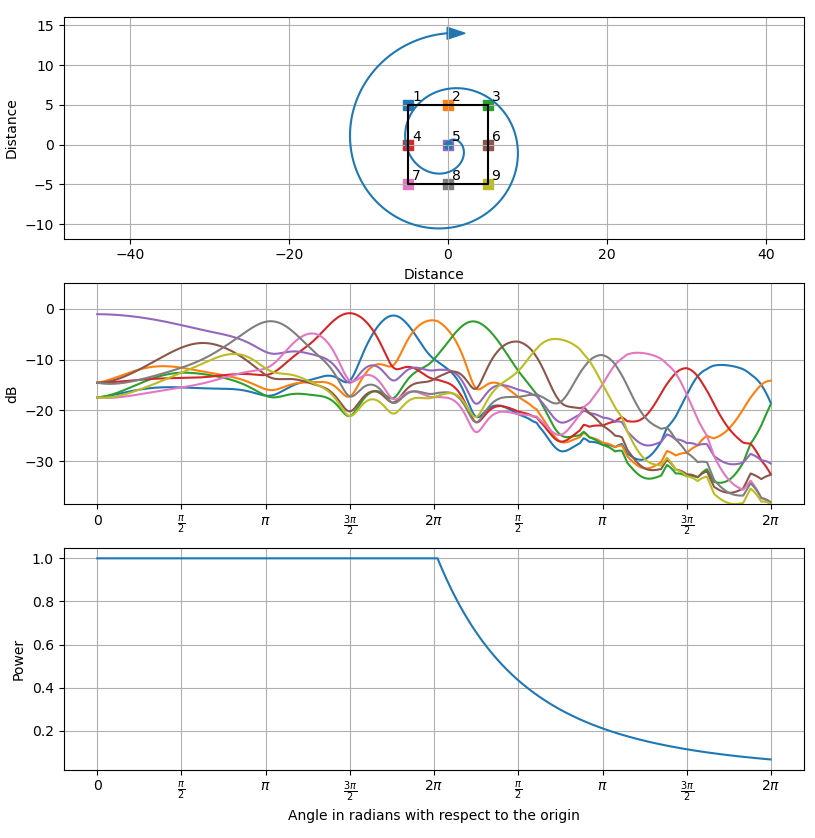}}
\caption{New DBAP. Rolloff = 6, $r_s = 1.073$. \emph{Top}: Plot of a 3x3 grid of loudspeakers with a line showing the movement of the source from the center of the field, extending in a spiral to a point outside the field. \emph{Middle}: Gains of the each speaker, color coded to correspond to the above figure. \emph{Bottom}: The power of the entire field.}
\label{new_gain_plots}
\end{figure}

The variable $p$ is the distance from a reference point in the field to the most distant speaker, $\max(d_{s}) = \max\{d_{s1}, ... , d_{sN} \} $, divided by the distance between the reference and the virtual source, $d_{rs}$, clipped to 1. In general, the reference point is best placed at the fields centroid, or geometic center, but this can also be modulated, such as in the case of tracking a listener moving through the field. In the case of Figs. \ref{original_gain_plots} and \ref{new_gain_plots}, the reference is placed at the origin.

This process effectively creates a circle around the reference with a radius equal to the distance between the source and most distant speaker, allowing for the power to equal 1 within this circle and to fall off at the rolloff as the virtual source moves away. In doing so, one can avoid calculation of the convex hull entirely with limited spatial distortion. This is especially valuable in a three-dimensional context where the calculation of a convex hull is difficult and resource-consuming. In Section \ref{conclusion}, however, the convex hull will continue to be used as concept in order to more accurately describe the position of the virtual source with respect to the loudspeakers.

% and
% \begin{equation}
%     r = \frac{ \max{(d_{is})} }{ d_{ir} }
% \end{equation}

\subsection{Biasing of loudspeakers for virtual sources far outside the hull}
The above method using $p$ creates satisfactory movement from a source moving from within to outside of the convex hull of loudspeakers. While this method works very well for sources that maintain relative ``closeness'' to the loudspeaker field, the problem described in \S 2.6 of \cite{dbap} of the source appearing to move toward the center as its distance grows remains.

While in practice this was not generally found to be an issue except in extreme circumstances, a method for biasing is presented here and works adequately in situations where a source cannot remain, for whatever reason, within a reasonable distance from the loudspeaker field. A biasing parameter, $b_i$ for the $i$th speaker, is added to \eqref{vi_orig} and \eqref{k_new} which become
\begin{equation}
  v_i = \frac{k w_i b_i}{d_i^a}
\end{equation}
\begin{equation}
  k = \frac{p^{2a}}{\sqrt{\sum_{i=1}^{N} \frac{b_i^2 w_i^2}{d_i^{2a}}}} \label{k_new_with_b}
\end{equation}
with $b_i$ defined as
\begin{equation}
  b_i = \Bigg( \frac{u_i}{u_{m}} \bigg( \frac{1}{p} - 1 \bigg) \Bigg)^2 + 1 \label{bi}
\end{equation}
\begin{equation}
  u_i = \big( d_i - \max(d) \big)_{normalized}^2 + \epsilon \label{ui}
\end{equation}
where $m$ is the index of the median distanced loudspeaker from the virtual source, $\max(d)$ is the loudspeaker furthest from the virtual source so that $\max(d) = \max\{d_1, ... , d_N\}$, and $\epsilon$ is a small value to avoid 0 gain in the most distant loudspeaker (typically set to $r_s/N$). Eq. \eqref{ui} is the normalized difference between the $i$th loudspeaker and the most distant loudspeaker, squared, plus $\epsilon$. One can see from \eqref{bi} and \eqref{ui} that $b_i$ increases when $d_i < d_m$ and decreases to a minimum of 1 when $d_i > d_m$. In other words, loudspeakers closer to the virtual source than $m$th loudspeaker have more ``weight'' than do those which are further from the source than loudspeaker $m$. This creates biasing without the need for projection and the effect is exaggarated as $d_i$ is larger or smaller than $d_m$. Note that $b_i$ is simply an additional weight parameter and can often be calculated in place of $w_i$ if no weighting is used.

% \subsection{Redefinition of variable $a$}
% As a matter of convenience, $a$ is redefined as
% \begin{equation}
%     a = \frac{20\log_{10} 2} {R}
% \end{equation}
% in order to produce a more natural result given the input rolloff, $R$. That is, when $a$ is defined as it appears in \cite{dbap}, the actual rolloff \emph{decreases} (amplitudes are attenuated less) as a function of distance when the value of $R$ is increased. Defining $a$ as above, a higher value for $R$ causes the rolloff to increase resulting in sharper amplitude curves.

% \section{Discussion}\label{discussion}
% The tradeoff in the DBAP algorithm, however configured, seems to be between that of steadily changing speaker amplitudes --- and therefore undulating power --- or steadily changing power --- and thus undulating amplitudes. In practice, it has been found that allowing the amplitudes to undulate and maintaining a smooth power curve creates fewer and less undesired artifacts than allowing the power to modulate, as in the case of ADBAP.

\section{Discussion}\label{conclusion}
While not guaranteed to be perfect, the methods proposed in this paper have been shown to provide convincing spatialization of a virtual source in practice in a wide variety of of loudspeaker layouts. These layouts further need not be, and in fact ought not be, conventional loudspeaker layouts such as 5.1, 7.1, 22.2, linear, rectangular, or circular layouts. There are superior spatial paradigms for layouts such as these and DBAP was never designed to exceed their performance in a conventional layout.

Although the methods presented in \S \ref{new_methods} somewhat complicate the caluation of gains when a virtual source is located within the convex hull, they create a superior spatial impression in all configurations and circumstances than \cite{dbap}. When a source is located outside of the convex hull, the calculations are simplified as there is no need to determine the orthogonal projection. Likewise, there is no need to calculate the convex hull at all since the gains of a source located outside the hull are not a function of projection.

In the case of the spatial blur, $r_s$, it was found that it is best to set this as a scalar of the average distance between the loudspeaker field centroid and each loudspeaker to enable easy adaptation to different layouts. This creates similar performance for different layouts while avoiding needless fiddling with parameters. The variable $r_s$ then becomes
\begin{equation}
    r_s = \Bigg( \frac{\sum^N_{i=1} d_{ic}}{N} \Bigg) r_{scalar}
\end{equation}
where $d_{ic}$ is the distance from the centroid to the $i$th loudspeaker. In practice, a value of $0.5 \geq r_{scalar} \geq 0.2$ performs well in most situations. In Figures \ref{original_gain_plots}, \ref{new_gain_plots}, and \ref{asym:orig_gain}--\ref{asym:new_gain}, $r_{scalar}$ is set to $0.2$.

Remaining difficulties in DBAP-like paradigms are related to the non-spherical emanation of sound from a traditional loudspeaker. That is, a listener in the same position in the loudspeaker field will necessarily have a different impression depending on the orientation of the loudspeakers themselves. Solving this would require the development of new hardware and is well beyond the scope of this article.

\begin{figure}
  % \centering{\includegraphics[width=0.5\textwidth]{figs/asym/field.png}}
  \centering{\includegraphics[width=0.5\textwidth]{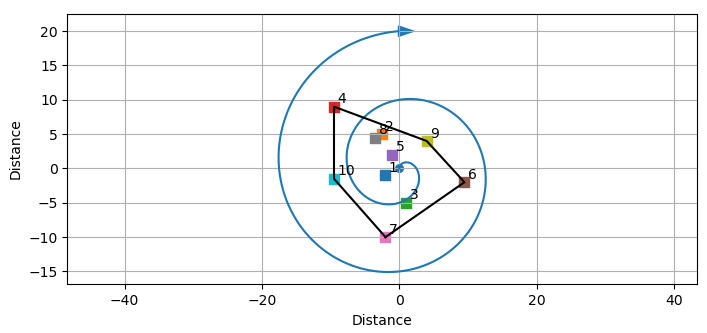}}
\caption{Speaker layout and movement of virtual source in the asymmetrical field.}
\label{asym:field}
\end{figure}

\begin{figure}
  % \centering{\includegraphics[width=0.5\textwidth]{figs/asym/orig_gain.png}}
  \centering{\includegraphics[width=0.5\textwidth]{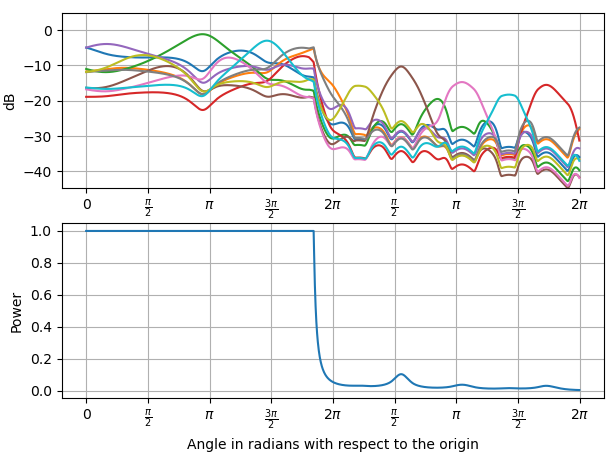}}
\caption{Gain curves for the original DBAP algorithm with projection. Note the very fast drop off in power once the virtual source exits the convex hull.}
\label{asym:orig_gain}
\end{figure}

\begin{figure}
  % \centering{\includegraphics[width=0.5\textwidth]{figs/asym/abs_gain.png}}
  \centering{\includegraphics[width=0.5\textwidth]{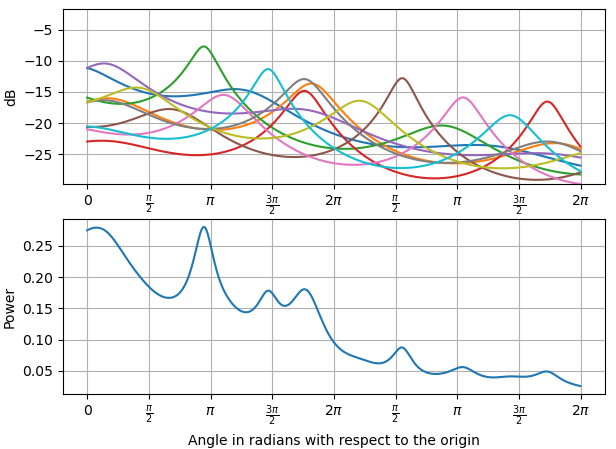}}
\caption{Gain curves for ADBAP. Although the behavior is much more what is expected, note the spike in power around the first $\pi$ when the virtual source closes in on the cluster of speakers 2 and 8.}
\label{asym:abs_gain}
\end{figure}

\begin{figure}
  % \centering{\includegraphics[width=0.5\textwidth]{figs/asym/new_gain.png}}
  \centering{\includegraphics[width=0.5\textwidth]{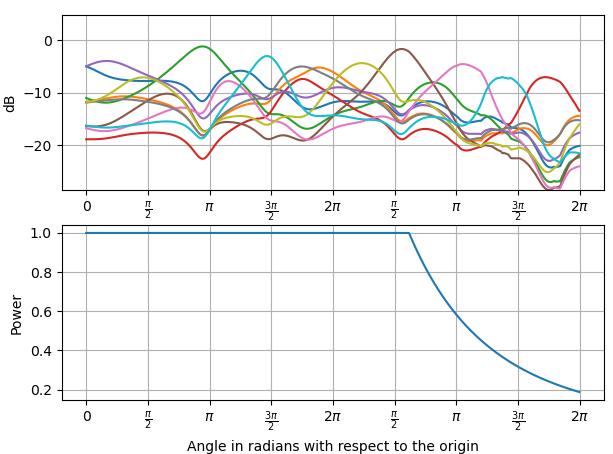}}
\caption{Gain curves for the version presetned in this paper.}
\label{asym:new_gain}
\end{figure}

\appendix
Figures \ref{asym:orig_gain}--\ref{asym:new_gain} show the performance of the original flavor of DBAP, ADBAP, and the version presented in this paper in an asymmetrical loudspeaker layout. The asymmetrical layout (Figure \ref{asym:field}) is ill-suited for analysis (hence its relegation to the Appendix) but is quite possible in a performance or installation context. For this reason, it is included here. The locations of the loudspeakers are (-2, -1), (-2.5, 5), (1, -5), (-9.5, 9), (-1, 2), (9.5, -2), (-2, -10), (-3.5, 4.5), (4, 4), and (-9.5, -1.5). Note that while the reference in calculations is set to (0, 0), the actual centroid of the loudspeaker field is (-1.55, 0.5). Even when the reference is not the centroid, the performance of the DBAP version presented here is far superior to that presented in \cite{dbap}.

\end{document}